\documentclass[%
 amsmath,amssymb,
 aps,
 twocolumn,
 reprint,
 floatfix,
]{revtex4-2}

\usepackage{float}
\usepackage{physics}
\usepackage{amsfonts}
\usepackage{xcolor}
\usepackage{enumerate}
\usepackage{lmodern}
\usepackage{upgreek}
\usepackage{dsfont}
\usepackage{bm}
\usepackage{mathtools}
\usepackage{enumitem}
\usepackage{dcolumn}
\usepackage{xargs}
\usepackage{mdframed}
\usepackage{etoolbox}
\apptocmd{\sloppy}{\hbadness 10000\relax}{}{}

\usepackage[utf8]{inputenc}
\usepackage[T1]{fontenc}

\DeclareMathOperator{\Span}{span}
\DeclareMathOperator{\STD}{STD}

\newcommand*\mean[1]{\overline{#1}}

\newcommand{\Krylov}{\mathcal{K}}
\newcommand{\Kc}{\Krylov_C}
\newcommand{\Kcmean}{\mean{\Krylov_C}}

\newcommand{\liouv}{\mathcal{L}}
\newcommand{\obs}{\mathcal{O}}
\newcommand{\sigmabn}{\sigma_\text{log}}

\begin{document}

\title{Assessing the saturation of Krylov complexity as a measure of chaos}

\author{Bernardo L. Español}
\author{Diego A. Wisniacki}%
\affiliation{Departamento de Física ``J. J. Giambiagi'' and IFIBA, FCEyN, Universidad de Buenos Aires, 1428 Buenos Aires, Argentina}


\begin{abstract}
	Krylov complexity is a novel approach to study how an operator spreads over a specific basis. Recently, it has been stated that this quantity has a long-time saturation that depends on the amount of chaos in the system.
	Since this quantity not only depends on the Hamiltonian but also on the chosen operator, in this work we study the level of generality of this hypothesis by studying how the saturation value varies in the integrability to chaos transition when different operators are expanded.
	To do this, we work with an Ising chain with a longitudinal-transverse magnetic field and compare the saturation of the Krylov complexity with the standard spectral measure of quantum chaos.
	Our numerical results show that the usefulness of this quantity as a predictor of the chaoticity is strongly dependent on the chosen operator.
\end{abstract}

\maketitle

\section{Introduction}

Throughout this century advances in technology have allowed us to control increasingly complex quantum systems. The devices supported by these systems are considered the near future for computation and information processing. Therefore, it is essential to understand their robustness and sensitivity to perturbations. Characterizing the complexity or chaos of quantum systems is a crucial step in the development of more efficient technologies.

The study of chaos in quantum systems began with a focus on their spectral properties. 
Under this scheme, a chaotic system is differentiated from an integrable one based on how similar its level statistics are to a random matrix system ~\cite{Bohigas1984,Gutzwiller1990,Haake1991,Hans2000}.
Later on, the approach turned slightly to a dynamical definition of quantum chaos.
Examples of this include the proposal of the Loschmidt echo, which measures the irreversibility of a system when it is disturbed, and the use of out-of-time ordered correlators (OTOCs), which measure the speed at which an operator spreads over the space of operators while evolving in time.

Another possible approach to measure the complexity of quantum evolutions uses the Krylov subspaces and the Krylov complexity ($\Krylov$-complexity from now on) \cite{Parker2019, Dymarsky2020, Rabinovici2021, H_rnedal2022, Bhattacharjee2022,Fan2022}.
In this subspace, any quantum system can be expressed as a one-particle hopping problem on a one-dimensional chain, with hopping coefficients being given by the so-called Lanczos coefficients.
Studying how much an operator spreads over this base is equivalent to studying how much the wave function of this fictional particle spreads over the chain. The $\Krylov$-complexity is a quantitative measure of this dynamic, and it has recently caused a great deal of activity within the community.
Many works study this metric for a wide variety of systems \cite{Barbon2019,Noh2021,Ballar_Trigueros2022,Bhattacharya2022, Bhattacharjee2022_2, Cao2021, Heveling2022, Rabinovici_sup2022, Rabinovici2022}.

Specifically, in Refs. \cite{Rabinovici_sup2022, Rabinovici2022} it was conjectured and observed in certain systems that the $\Krylov$-complexity has a lower late-time saturation value for integrable systems compared to chaotic ones.
The authors suggested that the distribution of Poissonian statistics of the separation of energy levels, characteristic of integrable systems, influences the Lanczos coefficients, resulting in more disperse values.
These irregularities provoke an effect of localization in the wave function, preventing it from exploring the whole chain efficiently, which translates to a lower saturation value of the $\Krylov$-complexity.
It is important to identify that this conjecture relies on two assumptions.
The first is that Poissonian statistics generate more erratic Lanczos coefficients and, the second is that this noise generates localization and a diminution of the saturation of the $\Krylov$-complexity.

The goal of this work is to explore the generality of these hypotheses. 
In particular, we study how the dispersion of Lanczos coefficients and the late-time saturation of the $\Krylov$-complexity depend on the amount of chaos in the system for different operators.
To accomplish this, we use an Ising spin chain with a longitudinal-transverse magnetic field; given a value for the transverse component, this system has an integrability-chaos transition varying the longitudinal component.
To explore the full space of local operators, we expand the Krylov basis for the sum of Pauli operators over each one of the spins. 
Our numerical results suggest that the dispersion of Lanczos coefficients is slightly correlated with chaos for the operators we studied. However, this relationship is not well defined and exhibits large fluctuations, indicating the need for larger systems or statistical ensembles of different operators to determine whether there is a defined systematicity.
Conversely, the saturation of the $\Krylov$-complexity tends to have a defined trend, but its correlation with chaos is strongly dependent on the operator chosen to construct the Krylov basis.

This work is organized as follows.
In Sec. \ref{sec:krylov} we briefly review the Krylov subspaces and $\Krylov$-complexity.
In Sec. \ref{sec:results} we begin by defining the dispersion measurement of the Lanczos coefficients and the parameters of the Hamiltonian we are going to work with.
Then, we generate the Krylov basis for different operators and for each one of them we study the behavior of the Lanczos coefficients and the saturation of the $\Krylov$-complexity by altering the level of chaos in the system.
We conclude in Sec. \ref{sec:discussion} with a summary and some final remarks.
In order to get a deeper understanding of the work, we have included appendixes with details of the Hamiltonian that we used to perform the numeric simulations, together with the study of its symmetries (Appendix \ref{sec:ising_transverse}),  the definition of the measure of chaos (Appendix \ref{sec:eta}) and a detailed explanation of the measure's normalization  (Appendix \ref{sec:normalization}). 

\section{Krylov subspace, Lanczos algorithm and $\Krylov$-complexity}\label{sec:krylov}

Krylov methods were first introduced in the mathematical literature as an efficient way to perform matrix exponentiation~\cite{Hochbruck1997, Parlett1998}. This is why it is a traditional approach to approximate the quantum evolution of systems with large Hilbert spaces.
Recently, Krylov subspaces began to be used in quantum many-body problems to study the behavior and complexity
developed by operators over time~\cite{Parker2019, Ruffinelli2022}.
In this section we show how to construct the basis of Krylov subspaces using the Lanczos algorithm, along with the definition of the $\Krylov$-complexity.

Given a Hamiltonian $H$, a Hermitian operator $\obs$, and the Liouvillian superoperator $\liouv = \comm{H}{\cdot}$, the Krylov subspace is defined as the minimum subspace of $\liouv$ that contains $\obs(t)$ at all times, that is,
\begin{equation}
	\Krylov = \Span\{|\obs),\,\liouv|\obs),\,\liouv^2|\obs), \dots\},
\end{equation}
where $|\obs)$ is the state representation of $\obs$ in the operator's Hilbert space.
Given an inner product $(\obs_1|\obs_2) = \Tr\,(\obs_1^\dag\,\obs_2)$, the iterative Lanczos algorithm can be used to generate an orthonormal basis of this subspace.
The exact form of this algorithm consists of the following steps:
\begin{itemize}
	\item[] Define auxiliary variables: \\ $b_0 = 0$, $~|\obs_{-1}) = 0$.
	\item[] Normalize the operator to expand:\\
		$|\obs_0) = |\obs) / \left(\obs|\obs\right)^{\frac{1}{2}}$.
	\item[] for $n = 1, 2, \cdots$, repeat:
		\begin{itemize}
			\item[] $|{\mathcal{U}}_n) = \liouv\,|\obs_{n-1}) - b_{n-1}\,|\obs_{n-2})$.
			\item[] $b_n = \left(\mathcal{U}_n|\mathcal{U}_n\right)^{\frac{1}{2}}$. If $b_n = 0$, stop.
			\item[] $|\mathcal{O}_n) = |\mathcal{U}_n) / b_n$
		\end{itemize}
\end{itemize}
Doing this, we obtain the orthonormal Krylov basis $\left\{|\obs_n)\right\}$ and the Lanczos coefficients $\left\{b_n\right\}$.
For a Hamiltonian system with a Hilbert space of finite dimension $D$, it can be proven that $n$ is upper bounded by $D^2 - D + 1$~\cite{Rabinovici2021}, such that the $|\obs_n)$ form a subspace of dimension $1 \leq K \leq D^2 - D + 1$.

Before proceeding, there are a couple of points to note.
Firstly, while this algorithm is useful for theoretical purposes, it is impractical to use on a computer due to the accumulation of error from floating point rounding in the inner products.
To address this issue, there are alternative implementations that allow for controlling this error, such as those described in~\cite{Parlett1998,Simon1984}.
The simplest of these, and the one used in this work, consists of orthonormalizing $\liouv|O_{n-1})$ with respect to all previous $|O_n)$ instead of just the preceding one.
Second, all implementations of this algorithm are unstable when the Hamiltonian matrix has degeneracies, so if the system has any symmetries, it is necessary to desymmetrize the Hamiltonian and study the operator in symmetry subspace.

In the Krylov basis, the Liouvillian has a tridiagonal form with zero diagonal
\begin{equation}\label{eq:louvillian_heisenberg_equiation}
	\liouv\,|\obs) = b_n |\obs_{n-1}) + b_{n+1}\,|\obs_{n+1}).
\end{equation}
Expanding $|\obs(t))$ in wave functions $\phi_n(t) = i^{-n} \left(\obs_n|\obs(t)\right)$, we obtain,
\begin{equation}
	|\obs(t)) = \sum_{n=0}^{K-1} i^n\,\phi_{n}(t)\,|\obs_n).
\end{equation}
Substituting in Eq. (\ref{eq:louvillian_heisenberg_equiation}) results in a discrete Schrödinger equation for the wave functions,
\begin{equation}
	\partial_t \phi_n(t) = b_n\phi_{n-1} - b_{n+1}\,\phi_{n+1},
\end{equation}
with $b_0 = \phi_{-1} = 0$.
Note that this differential equation is the same as the one for a tight-binding problem in one dimension, with hopping coefficients $b_n$ between sites $\phi_n$ and $\phi_{n-1}$ on the chain.
Since $|\obs(t=0)) = |\obs_0)$, at $t=0$ the wave function is fully localized at the first site $\phi_n(0)=\delta_{0n}$.

To study how information spreads over the chain after the initial time, the following notion of complexity is defined:
\begin{equation}
	\Kc(t) = \sum_{n=0}^{K-1} n\,\abs{\phi_n(t)}^2.
\end{equation}
This time-dependent quantity is called $\Krylov$-complexity and it is simply the mean value of the position of the $\phi_n(t)$ in the chain.

As we interested in its late-time value, we define
\begin{equation}
	\Kcmean = \sum_{n=0}^{K-1} n\,\mean{\abs{\phi_n(t>\tau)}^2}.
\end{equation}
with $\tau$ being the time at which complexity saturates.

\section{From integrability to chaos through Lanczos Coefficients}\label{sec:results}

In this section we study how the dispersion of the Lanczos coefficients and the saturation of the $\Krylov$-complexity depend on the chaoticity of the system.
The dispersion of an ordered set of values can be defined in different ways.
To facilitate the discussion and comparison with previous studies \cite{Rabinovici_sup2022, Rabinovici2022},  we use as a measure of dispersion the standard deviation of the logarithmic difference between the successive values of the chain,
\[
	\sigmabn = \STD\{\log(b_n) - \log(b_{n+1})\} = \STD\{\log(b_n/b_{n+1})\},
\]
where the $b_n$ are the Lanczos coefficients.

As a model, we use a spin chain with a longitudinal-transverse magnetic field.
The Hamiltonian of this system and the study of its symmetries can be found in Appendix \ref{sec:ising_transverse}.
For all Lanczos coefficient calculations, we use $L=6$ (resulting in a Krylov space with dimension $K = 1261$), $J=1$ and $h_x=1$ and study the chaos-integrability transition by varying $h_z$.
We found similar results using $L=4$ ($K = 91$), $L=5$ ($K = 381$), and $L=7$ ($K = 5113$) (in the latter case, only a limited number of points were calculated due to computational cost).
The operators we use are,
\begin{equation}
	S^\alpha_\text{T} = \sum_{k = 1}^L \frac{1}{2}\,\sigma^{\alpha}_k
\end{equation}
where $\sigma^\alpha_k$ is the Pauli operator at site $k$ with direction $\alpha = \{x, y, z\}$.
In addition, we compare these results with those obtained from a random Gaussian, real, traceless operator, which is a strongly entangled operator over the entire chain. 

\begin{figure}[ht]
	\centering
	\includegraphics[width=\linewidth]{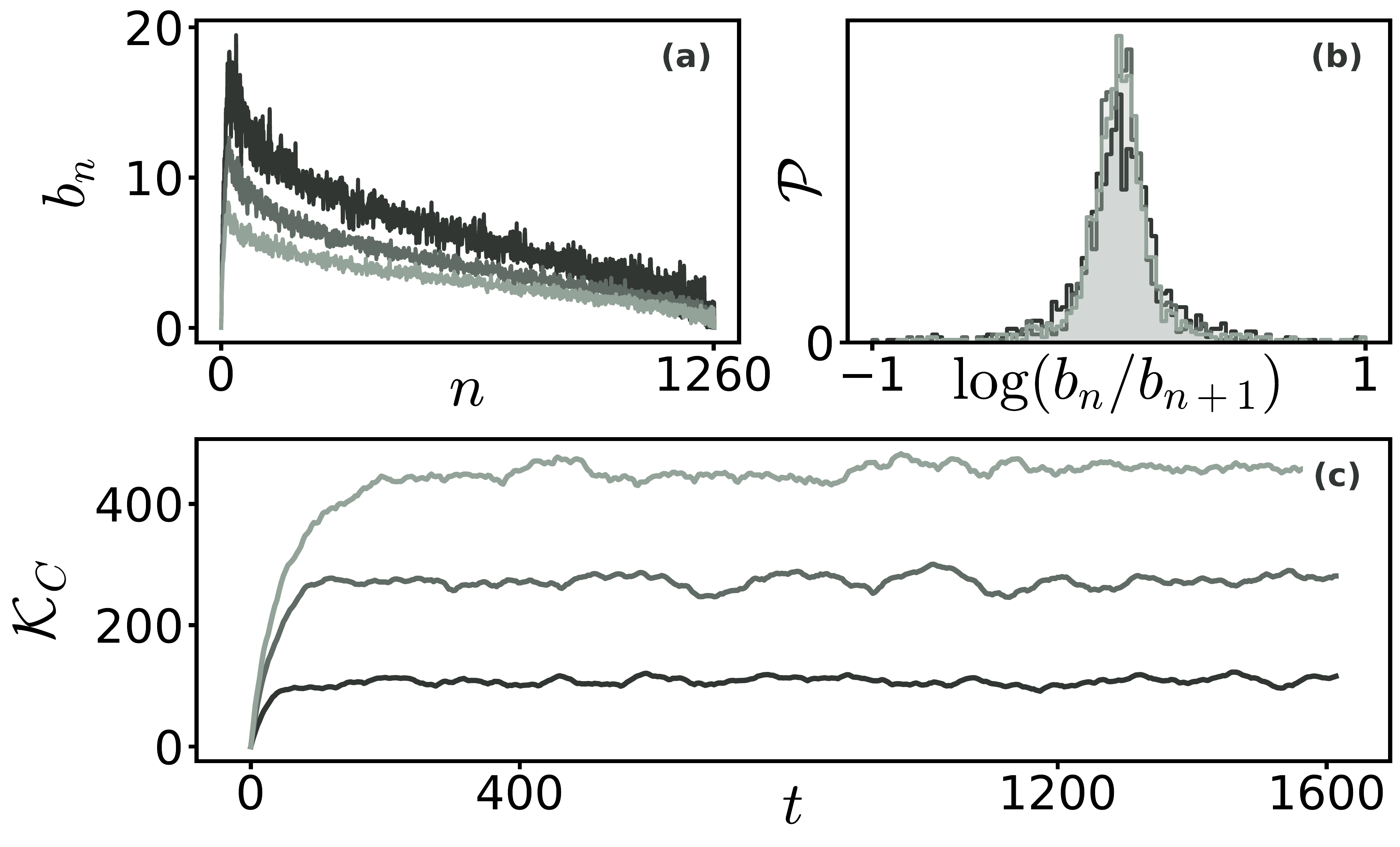}
	\caption{
			Krylov expansion of $\obs=S^z_\text{T}$.
			From lightest to darkest, the curves represent $h_z=0.2$ and $\eta \simeq 1$, $h_z = 1.35$ and $\eta \simeq 0.7$, $h_z=2.5$ and $\eta \simeq 0.1$.
			(a) Lanczos sequence as a function of $n$.
			(b) Distribution of the logarithmic difference between elements of the sequences. The mean of these distributions in the three cases is practically zero, and their dispersion is $\sigmabn \simeq 0.32$ for $\eta \simeq 1$, $\sigmabn \simeq 0.45$ for $\eta \simeq 0.7$, and $\sigmabn \simeq 0.6$ for $\eta \simeq 0.1$.
			(c) Plot of the $\Krylov$-complexity as a function of time for the three calculated sequences.
			The late-time saturations values are $\Kcmean=459\pm10$ for $\eta\simeq1$, $\Kcmean=274\pm11$ for $\eta\simeq0.7$, and $\Kcmean=108\pm6$ for $\eta\simeq0.1$
    }
	\label{fig:transverse_sz_3p}
\end{figure}

Let us start by taking $\obs = S^z_\text{T}$.
In Fig. \ref{fig:transverse_sz_3p} (a) we show the Lanczos coefficients for three different values of $h_z$, one in the integrable regime ($h_z=2.5$), another in the chaotic regime ($h_z=0.2$), and a third at a point between the two ($h_z=1.35$).
The dispersion of these coefficients in each regime is shown in Fig. \ref{fig:transverse_sz_3p}(b) and the $\Krylov$-complexity is plotted as a function of time in Fig. \ref{fig:transverse_sz_3p}(c).
We quantify the chaoticity of the system using the standard measure $\eta$ , which is a spectral measure 
that is equal to $0$ if the system is integrable and equal to $1$ if it is chaotic.
The $\eta$ is obtained from the distribution of $\min(r_n,1/r_n)$, where $r_n$ is the ratio between the two nearest neighbor spacings of a given level ($r_n = s_n/s_{n-1}$).
In Appendix \ref{sec:eta} we give more details about how this quantity is calculated. The $\eta$ was computed in a chain with $L=13$.
For the three cases presented in Fig. \ref{fig:transverse_sz_3p} there seems to be a relationship between the level of chaos present in the system $\eta$, $\sigmabn$ and $\Kcmean$.
Specifically, as the system becomes more chaotic, the dispersion of $b_n$ decreases and the saturation point of the $\Krylov$-complexity increases.

\begin{figure}[ht]
	\centering
	\includegraphics[width=\linewidth]{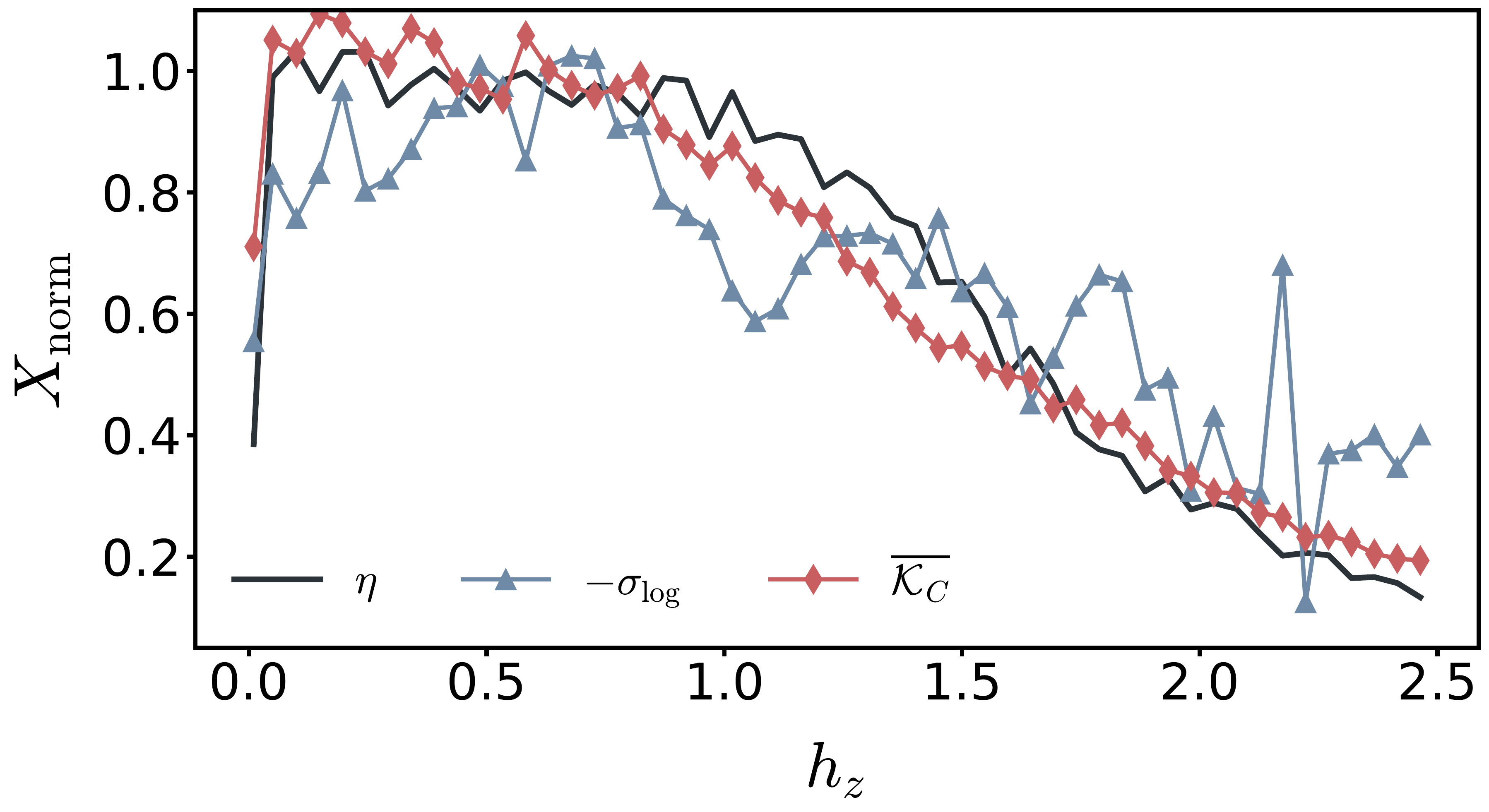}
	\caption{
		Comparison between the degree of chaos present in the system (indicated by $\eta$), the dispersion of the Lanczos sequences $\sigmabn$, and the late-time saturation value of the $\Krylov$-complexity $\Kcmean$, as a function of $h_z$, for $\obs = S^z_\text{T}$. All quantities were normalized in order to compare them on the same scale.
	}
	\label{fig:all_sz_together}
\end{figure}

To study the integrability to chaos transition in more detail in Fig. \ref{fig:all_sz_together} we show $\sigmabn$ and $\Kcmean$ compared to $\eta$ varying $h_z$ between $0.01$ and $2.5$.
These quantities were normalized so that they can be compared on the same scale.
The normalization that we use is discussed in Appendix \ref{sec:normalization}.
Since we expect $\sigmabn$ to decrease as the chaos parameter increases, we take $-\sigmabn$ to make all the quantities involved follow the same systematicity (all increase as a function of chaos).
We see that in the limits $h_z \ll h_x$ and $h_z \gg h_x$ the system is fundamentally integrable, and that for values of $h_z \sim h_x$ there is a transition indicated by the value of $\eta$ at each point.
Expanding the operator $S^z_\text{T}$, we find that the dispersions of the both $b_n$ and $\Kcmean$ are sensitive to this transition, exhibiting a behavior similar to $\eta$.
These results are consistent with those discussed Refs.~\cite{Rabinovici_sup2022,Rabinovici2022}.

\begin{figure}[ht]
	\centering
	\includegraphics[width=\linewidth]{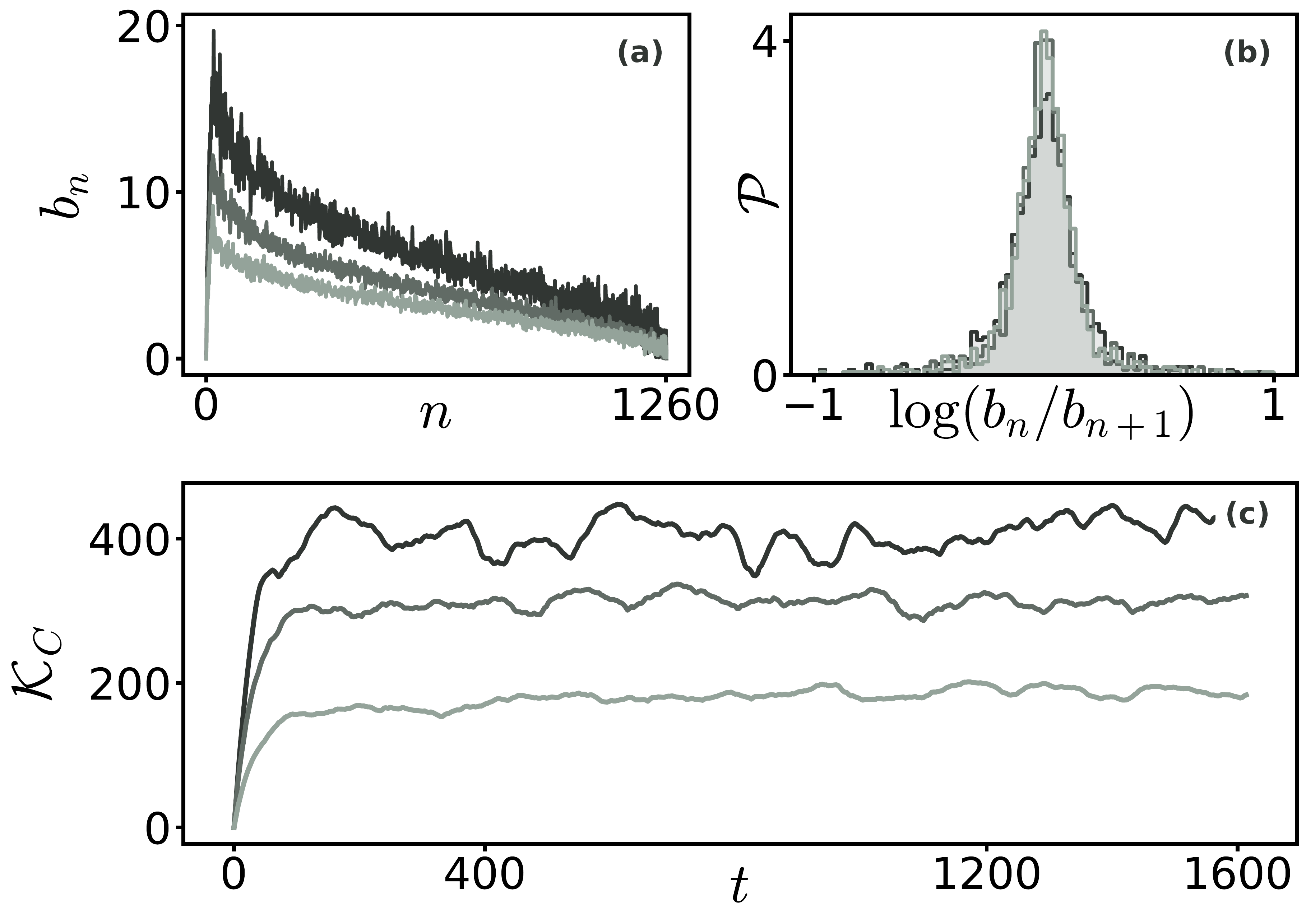}
	\caption{
		Krylov expansion of $\obs=S^x_\text{T}$.
		The system parameters and panel distribution are the same as in Fig. \ref{fig:transverse_sz_3p}.
		The dispersion of $\log(b_n/b_{n+1})$ for this operator is $\sigmabn \simeq 0.32$ for $\eta \simeq 1$, $\sigmabn \simeq 0.44$ for $\eta \simeq 0.7$, and $\sigmabn \simeq 0.6$ for $\eta \simeq 0.1$; with practically zero mean in all cases.
		The saturation of the $\Krylov$-complexity is $\Kcmean=188\pm7$ for $\eta\simeq1$, $\Kcmean=316\pm13$ for $\eta\simeq0.7$, and $\Kcmean=417\pm25$ for $\eta\simeq0.1$.
	}
	\label{fig:transverse_sx_3p}
\end{figure}

\begin{figure}[ht]
	\centering
	\includegraphics[width=\linewidth]{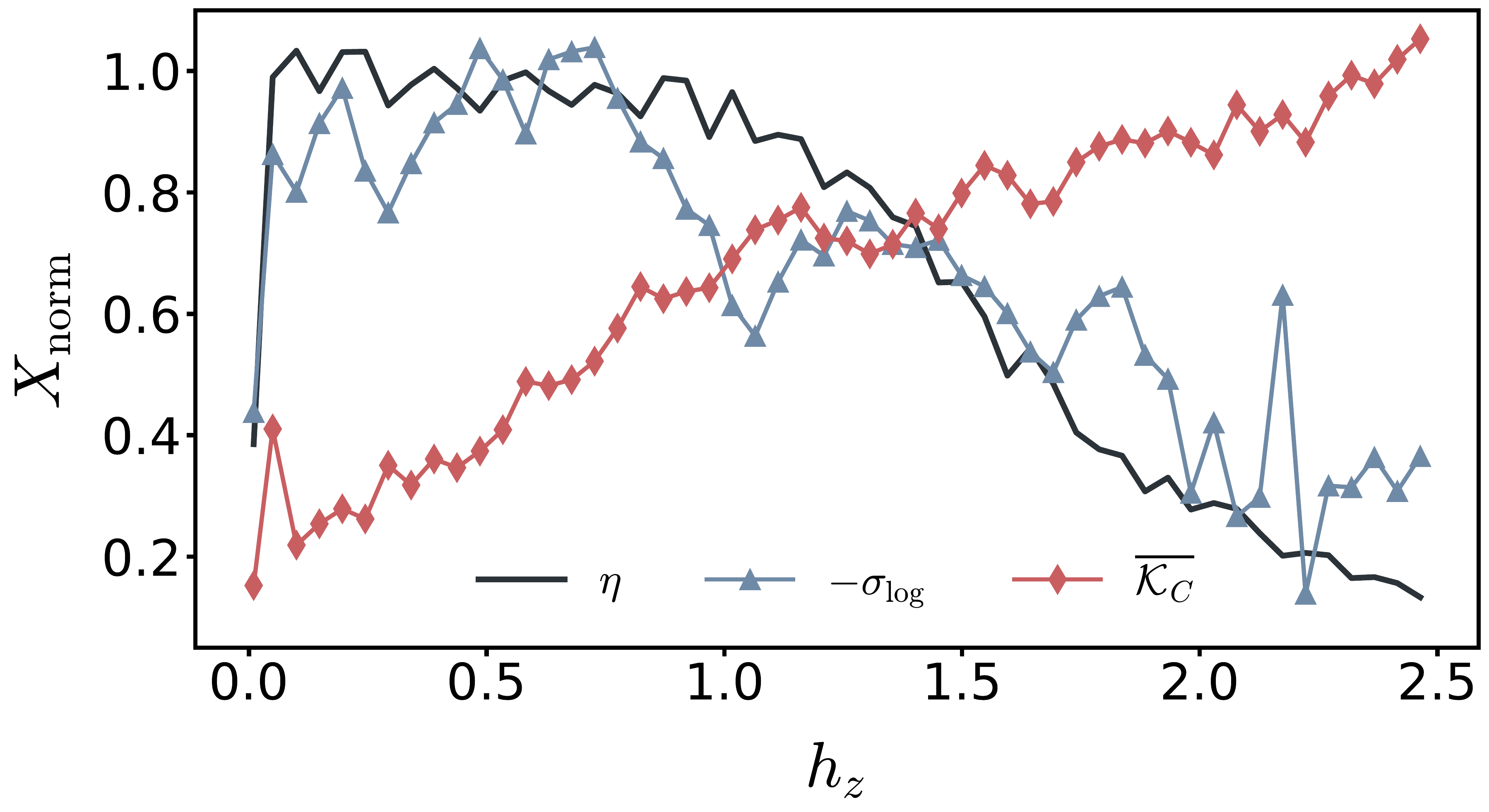}
	\caption{
		Comparison between the degree of chaos present in the system (indicated by $\eta$), the dispersion of the Lanczos sequences $\sigmabn$, and the late-time saturation value of the $\Krylov$-complexity $\Kcmean$, as a function of $h_z$, for $\obs = S^x_\text{T}$. All quantities were normalized in order to compare them on the same scale.
	}
	\label{fig:all_sx_together}
\end{figure}

Now we consider the case $\obs = S^x_\text{T}$. Fig. \ref{fig:transverse_sx_3p} was constructed in the same way as Fig.  \ref{fig:transverse_sz_3p}, but with the Krylov basis generated using this operator.
In this case, the saturation of the $\Krylov$-complexity has a notably different pattern. Comparing these two figures, even though the Lanczos sequences are practically the same ( in both average value and dispersion), the $\Krylov$-complexity for $S^x_\text{T}$ exhibits behavior opposite to that seen for $S^z_\text{T}$.
In this case, when the system is chaotic, the $\Krylov$-complexity saturates at a lower value than when it is integrable. 
In Fig. \ref{fig:all_sx_together}, this behavior can be seen extended for the same values of $h_z$ as in Fig. \ref{fig:all_sz_together}.

\begin{figure}[ht]
	\centering
	\includegraphics[width=\linewidth]{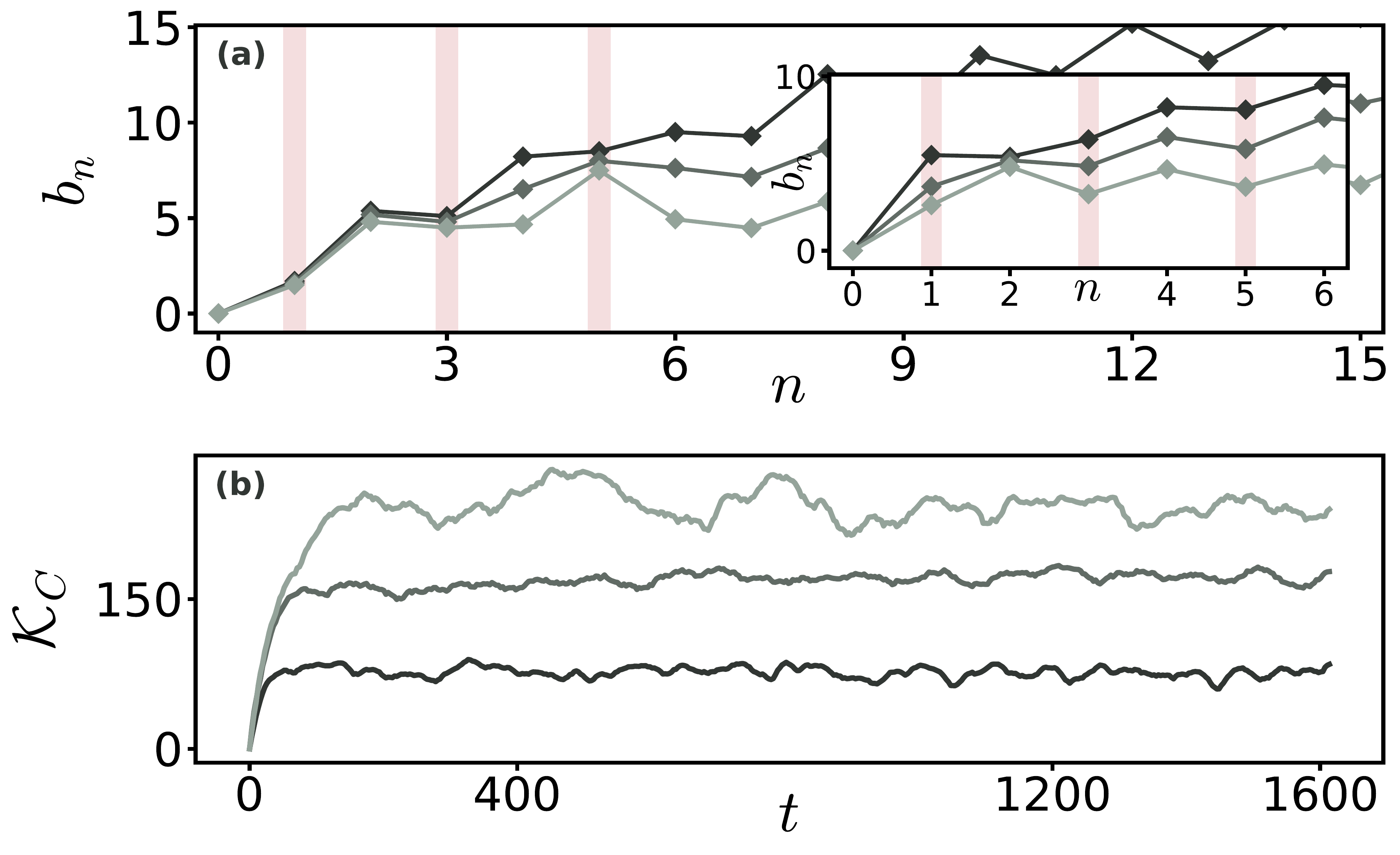}
	\caption{
			The inset in (a) shows a close-up of the first Lanczos coefficients shown in Fig. \ref{fig:transverse_sx_3p}(c).
			(a) Same coefficients but handpicking the values of $b_1$, $b_3$, and $b_5$.
			(b) Plot of the $\Krylov$-complexity for the modified chain.
	}
	\label{fig:transverse_sx_c_b135}
\end{figure}

These results show not only that $\Kcmean$ depends on the operator, but also that there is no direct relationship between the dispersion of $b_n$ and $\Kcmean$.
We can see this same idea by handpicking some of the first $b_n$.
In Fig. \ref{fig:transverse_sx_c_b135} we observe that by changing only the values of $b_1, b_3$ and $b_5$ of the $S^x_\text{T}$ sequence we can completely modify the dynamics of the $\Krylov$-complexity, even to the point of inverting the original relationship (now the configuration that saturates at a higher value is the chaotic regime, as when we expand $\obs=S^z_\text{T}$).
Although it is not clear which values of the Lanczos coefficients are the ones that determine the general behavior of the $\Krylov$-complexity, in our numerical tests we find that $\Kcmean$ is very sensitive to the first few values of $b_n$, and that it can be changed qualitatively by changing only a few coefficients; without affecting the statistics of the entire sequence.
These results are consistent with the idea that the first elements of the basis are the ones that dominate the dynamics of the operator \cite{Parker2019, Noh2021}.
In particular, in Ref. \cite{Noh2021} it was shown for the integrable regime of the same system ($h_x= 0$ and $h_z = 1$) that the first Lanczos coefficients grow linearly when expanding the one-body operator $\sigma_0^z$, while those of $\sigma_0^x$ rapidly converge to a constant value.
This observation is compatible with our result, but a more in-depth study is necessary to reach a robust conclusion.

\begin{figure}[ht]
	\centering
	\includegraphics[width=\linewidth]{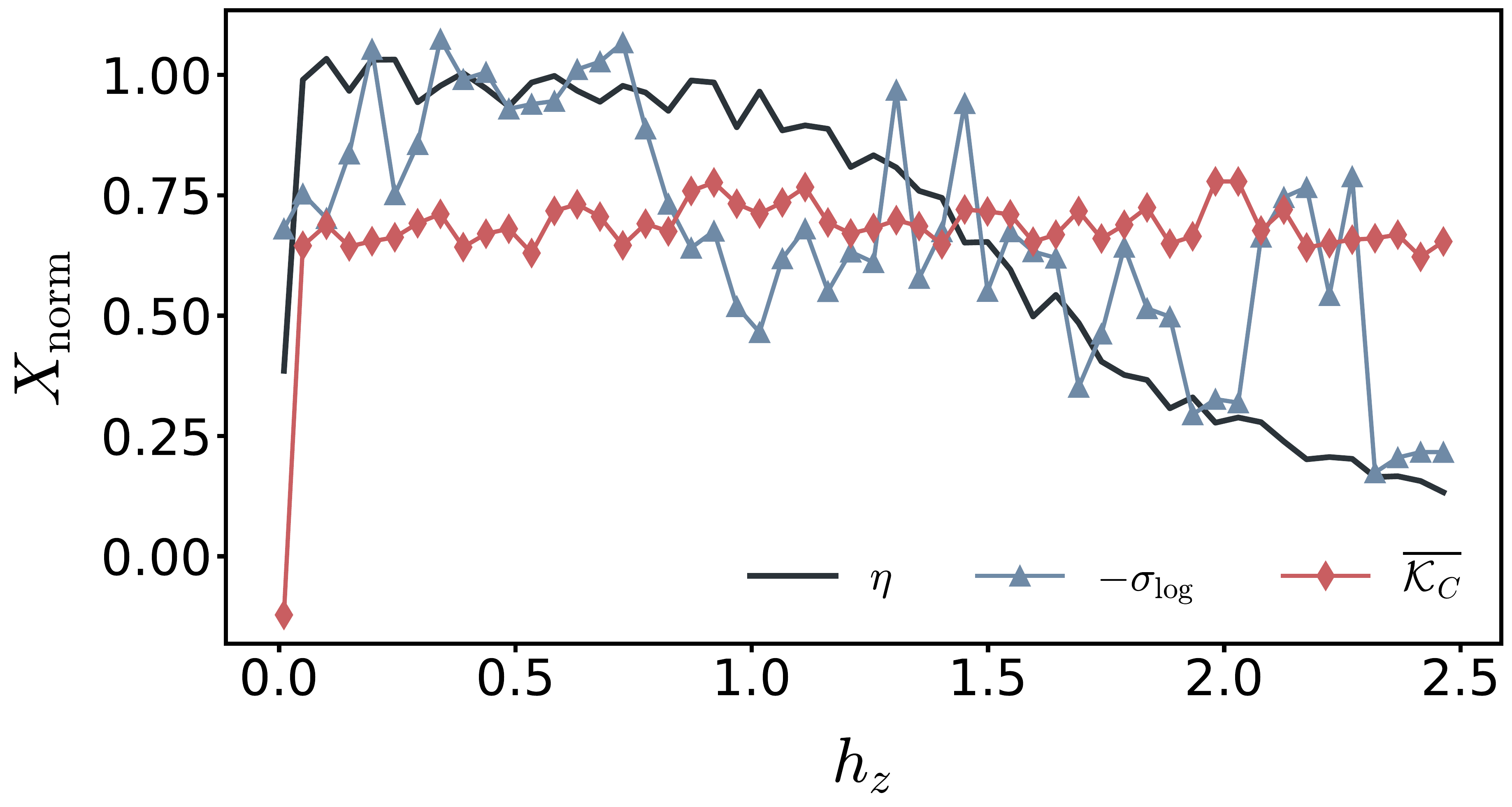}
	\caption{
		Comparison between the degree of chaos present in the system (indicated by $\eta$), the dispersion of the Lanczos sequences $\sigmabn$, and the late-time saturation value of the $\Krylov$-complexity $\Kcmean$, as a function of $h_z$, for a random Gaussian, real, traceless operator. All quantities were normalized in order to compare them on the same scale.
	}
	\label{fig:all_random_together}
\end{figure}

Finally, in Fig. \ref{fig:all_random_together} we show the transition when using the Krylov basis generated for a random-Gaussian, real, traceless operator.
We see that for this case the saturation of the $\Krylov$-complexity is not sensitive to the amount of chaos present in the system. When we expand $S^y_\text{T}$, we obtain a similar result.

It is worth mentioning that while all operators shown in this work have full lattice support, our calculations indicate comparable results when the initial operators have support only on a limited number of lattice sites.
As an example, we see that by expanding $S^z_T$ we get the same chaos dependence shown in Ref. \cite{Rabinovici2022} for $S_i^z + S_{L-i+1}^z$ ($i$ is chosen to be near the center of the chain).
However, taking  operators that combine different directions of Pauli operators, such as $S_i^z + S_{L-i+1}^x$, results in a chaotic dependence closer to that exhibited by $S_y^T$ or a random operator (not shown).

\section{Discussion}\label{sec:discussion}
In this work we have discussed the feasibility of using the saturation of the Krylov complexity as a measure of the chaos present in a system.
To do this, we worked with a Hamiltonian system with an integrability-chaos transition and compared this quantity with the usual spectral chaos measure.
We showed that the correlation between the $\Krylov$-complexity saturation and the chaoticity of the system strongly depends on the operator chosen to construct the Krylov basis.
In particular, we showed that there are operators for which this quantity correlates (Fig. \ref{fig:all_sz_together}), others for which it anti-correlates (Fig. \ref{fig:all_sx_together}) and others for which there is no clear systematicity (Fig. \ref{fig:all_random_together}).
This work provides the possibility to understand what the characteristics of the operators should be for the $\Kcmean$ to be a good predictor.
Some progress has been made in~\cite{Rabinovici2022}, where it is shown that operators must have zero trace, but clearly this is not enough.
Even in the cases where a relationship is observed, there are no robust theoretical or numerical results showing that there must be a causal relationship.
These results serve as a counter example to the expected relationship between the dispersion of the Lanczos coefficients and the saturation of the $\Krylov$-complexity.
In all the cases shown here, the dispersion of the Lanczos coefficients has a similar systematicity, while the $\Kcmean$ changes qualitatively.

It is also worth mentioning that this type of problem has not been observed in OTOCS, in fact, it has been shown that the long-time regime of the OTOCs is a good predictor of the integrability-chaos transition as long as the operators are local, even without desymmetrizing the system~\cite{Garc_a_Mata2018,Fortes2019}.
We consider that understanding the difference in this transition is essential for studying the relationship between Krylov complexity and OTOCs.

\appendix

\section{Ising spin chain in a transverse magnetic field}\label{sec:ising_transverse}

The system we use is an Ising model with nearest-neighbor interactions in the $z$ direction and a transverse magnetic field on the $(x, z)$ plane, whose Hamiltonian is given by
\begin{equation}
	\hat H = \sum_{k=1}^{L} \left(h_x {\hat \sigma^x}_k + h_z {\hat \sigma^z}_k\right) - J\, \sum_{k=1}^{L-1} {\hat \sigma^z}_k {\hat \sigma^z}_{k+1},
\end{equation}
where $L$ is the number of spin-$\frac{1}{2}$ sites in the chain, $\sigma^j_k$ is the Pauli operator at site $k = {1, 2, \dots, L}$ in the $x, y $ and $z$ directions, $h_x$ and $h_z$ are the components of the magnetic field (transverse and longitudinal, respectively), and $J$ is the nearest-neighbor coupling.

To expand an operator in the Krylov space, it is necessary to work in a symmetry subspace.
This system is invariant under reflection with respect to the center of the chain; the parity operator commutes with the Hamiltonian, allowing it to be decomposed into parity-even and parity-odd subspaces of dimensions $D = D^\text{even} + D^\text{odd}$, $D^\text{even/odd} \simeq D/2$.
In this work, we always operate in the positive parity subspace.

While this model is integrable in the limit of $h_x \gg h_z$ and $h_z \ll h_x$, it exhibits quantum chaos when the longitudinal and the transverse field are of comparable strength \cite{Mirkin_IOP2021}.

\section{Spectral characterization of quantum chaos}\label{sec:eta}

In order to have a notion of the chaos present in the system when varying the parameters of the Hamiltonian, we use a spectral measure.
Under this scheme, a quantum system is considered chaotic if the distances between the consecutive levels of the Hamiltonian follow a Wigner-Dyson distribution, and it is considered integrable if it is Poissonian.
Given a Hamiltonian $H$ in a symmetry subspace, if we group the energy levels into an ordered set $e_n$, we can define the nearest neighbor spacing as $s_n = e_{n+1}-e_n$.
To quantitatively measure the distance between the distribution of a given system and a perfectly chaotic or integrable one, it is common to define the indicator $\tilde{r}_n = \min(r_n, 1/r_n)$, where $r_n = s_n / s_{n-1}$. 
The average value of this indicator is minimum if the distribution of the $s_n$ is Poissonian ($\tilde{r}_\text{P}$) and maximum if it is a Wigner-Dyson  distribution ($\tilde{r}_\text{WD}$), so we can normalize this quantity as
\begin{equation}
\eta = \frac{\mean{\tilde{r}_n} - \tilde{r}_\text{P}}{\tilde{r}_\text{WD}-\tilde{r}_\text{P}}.
\end{equation}
This allows us to identify a system as integrable if $\eta \simeq 0$ and chaotic if $\eta \simeq 1$.

\section{About the normalization of the quantities}\label{sec:normalization}
The spectral indicator $\eta$ defined in Appendix \ref{sec:eta} is normalized so that its values are restricted to the $[0, 1]$ interval.
However, the dispersion of Lanczos coefficients $\sigmabn$ and the saturation of the $\Krylov$-complexity $\Kcmean$ are not.
For a better comparison, it is necessary to define a general normalization criterion.
The $\mean{r}$ normalization is natural, since we know its value when the system is integrable and when it is chaotic, instead, for $\sigmabn$ and $\Kcmean$ this is precisely what we want to study.

\begin{figure}[ht]
	\centering
	\includegraphics[width=0.95\linewidth]{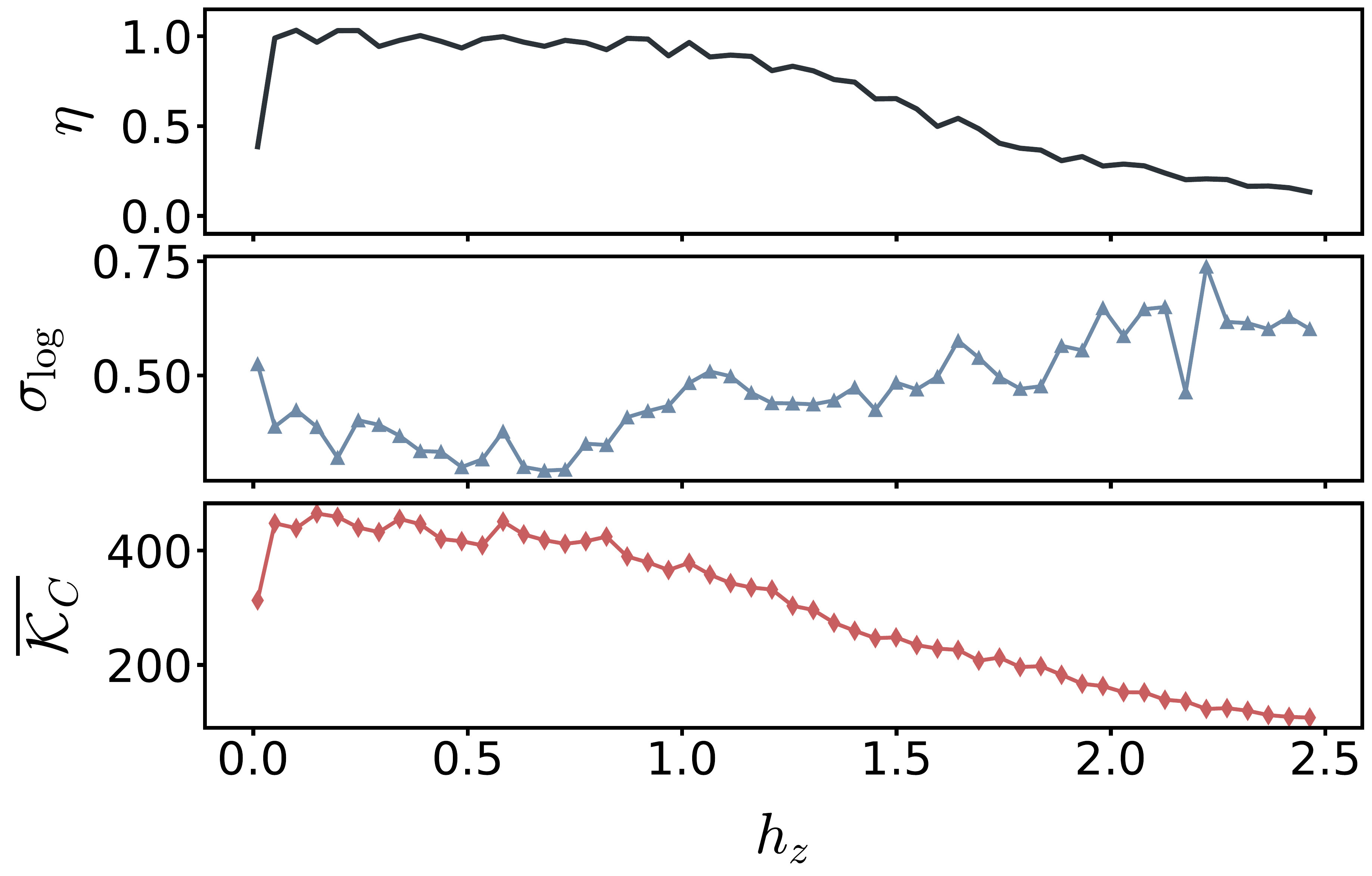}
	\caption{
		Same values as in Fig. \ref{fig:all_sz_together}, without normalization.
	}
	\label{fig:all_sz_separated}
\end{figure}

One possible option is mapping all the values to the $[0, 1]$ interval.
This consists of, given a sequence of values $X$, applying the transformation
\begin{equation}
	X^{(0, 1)}_\text{norm} = \frac{X - \min X}{\max X - \min X}.
\end{equation}
This normalization has the advantage of confining all magnitudes on the same interval, but it has two main problems:
i) $\eta^{(0,1)}_\text{norm}$ loses its intuitive normalization and
ii) if $X$ is noisy, $\min X$ will be poorly estimated, and this error will spreads to all points of $X^{(0, 1)}_\text{norm}$.

To handle the first problem we can preserve $\eta$ in its original interval and map the other curves onto it, by taking $X' = X\,\frac{\max \eta - \min \eta}{\max X - \min X}$.
To solve the second problem, the idea is to choose a displacement that does not rely on a single point, which will reduce the amount of error that is propagated.
\begin{figure}[ht]
	\centering
	\includegraphics[width=0.95\linewidth]{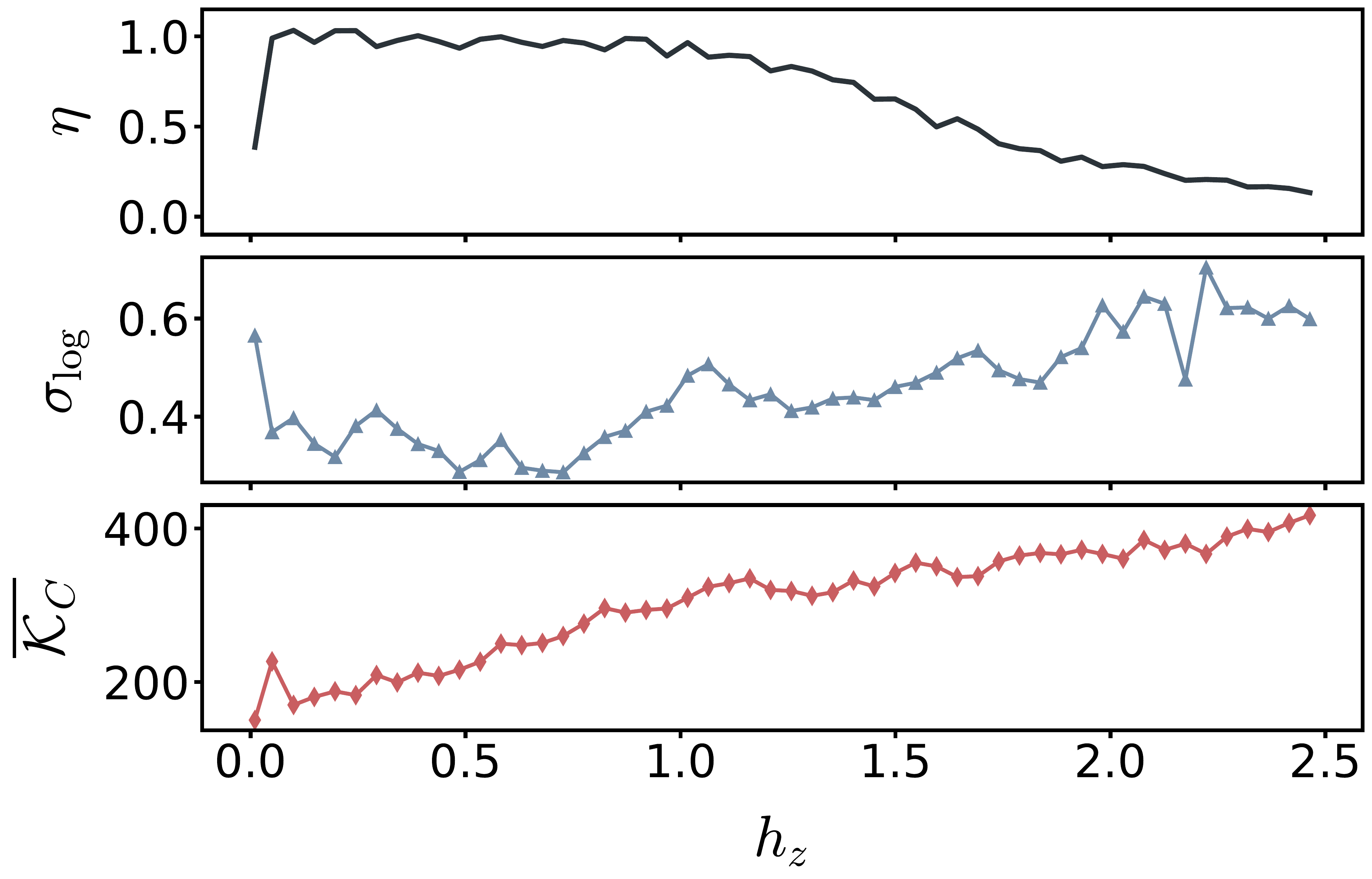}
	\caption{
		Same values as in Fig. \ref{fig:all_sx_together}, without normalization.
	}
	\label{fig:all_sx_separated}
\end{figure}
In this work we use
\begin{equation}
	\begin{aligned}
		X_\text{norm} = d_\text{min} (X', \eta) - X'\qc\\
		d_\text{min}(X', \eta)	= \min_\alpha \norm{(X'-\alpha) - \eta}
	\end{aligned}
\end{equation}
where $\norm{\cdot}$ is the Euclidean norm. 
This way $X'$ is displaced in order to minimize its Euclidean distance to $\eta$.

It is worth mentioning that the normalization of these quantities is arbitrary and only serves to scale their magnitudes so that they can be easily compared in the same plot. The conclusions of this work do not depend on this normalization. To emphasize this point and facilitate the reproducibility of the results, we show $\eta$, $\sigmabn$ and $\Kcmean$ without normalization for the $S^z_\text{T}$ and $S^x_\text{T}$ operators in Figs. \ref{fig:all_sz_separated} and \ref{fig:all_sx_separated} respectively.

\bibliography{bib} 

\end{document}